\documentclass{article}

\usepackage{arxiv}

\usepackage[utf8]{inputenc} 
\usepackage[T1]{fontenc}    
\usepackage{hyperref}       
\usepackage{url}            
\usepackage{booktabs}       
\usepackage{amsfonts}       
\usepackage{nicefrac}       
\usepackage{microtype}      
\usepackage{lipsum}
\usepackage{graphicx}
\usepackage{amsmath}
\usepackage{array}
\graphicspath{ {./images/} }

\title{Skill-Based Autonomous Agents for Material Creep Database Construction}

\author{
  Yue Wu$^{a,*}$ \\ 
  $^{a}$Materials Genome Institute\\
  Shanghai University\\
  Shanghai, China 200444 \\
  \texttt{repeatsongyue@gmail.com} \\
  \And
  Tianhao Su$^{a,*}$ \\
  $^{a}$Materials Genome Institute\\
  Shanghai University\\
  Shanghai, China 200444 \\
  \texttt{thsu0407@gmail.com} \\
  \And
  Shunbo Hu$^{a,b,\dagger}$ \\
  $^{a}$Materials Genome Institute, Shanghai University\\
  $^{b}$Institute for the Conservation of Cultural Heritage,\\
  Institute for Quantum Science and Technology,\\
  Key Laboratory of Silicate Cultural Relics Conservation\\
  (Ministry of Education),\\
  Shanghai University\\
  Shanghai, China 200444 \\
  \texttt{shunbohu@shu.edu.cn} \\
  \AND
  Deng Pan$^{a,\dagger}$\\
  $^{a}$Materials Genome Institute\\
  Shanghai University\\
  Shanghai, China 200444 \\
  \texttt{DPan\_MGI@shu.edu.cn} \\
}
\date{}

\begin{document}
\maketitle 
\begingroup 
    \renewcommand{\thefootnote}{\fnsymbol{footnote}}
    \footnotetext[1]{These authors contributed equally to this work.}
    \footnotetext[2]{Corresponding author.}
\endgroup

\date{\vspace{-4ex}}
\maketitle
\thispagestyle{empty}
\thispagestyle{plain} 
\begin{abstract}
The advancement of data-driven materials science is currently constrained by a fundamental bottleneck: the vast majority of historical experimental data remains locked within the unstructured text and rasterized figures of legacy scientific literature. Manual curation of this knowledge is prohibitively labor-intensive and prone to human error. To address this challenge, we introduce an autonomous, agent-based framework powered by Large Language Models (LLMs) designed to excavate high-fidelity datasets from scientific PDFs without human intervention. By deploying a modular "skill-based" architecture, the agent orchestrates complex cognitive tasks—including semantic filtering, multi-modal information extraction, and physics-informed validation. We demonstrate the efficacy of this framework by constructing a physically self-consistent database for material creep mechanics, a domain characterized by complex graphical trajectories and heterogeneous constitutive models. Applying the pipeline to 243 publications, the agent achieved a verified extraction success rate exceeding 90\% for graphical data digitization. Crucially, we introduce a cross-modal verification protocol, demonstrating that the agent can autonomously align visually extracted data points with textually extracted constitutive parameters ($R^2 > 0.99$), ensuring the physical self-consistency of the database. This work not only provides a critical resource for investigating time-dependent deformation across diverse material systems but also establishes a scalable paradigm for autonomous knowledge acquisition, paving the way for the next generation of self-driving laboratories.
\end{abstract}


\section{Introduction}
The landscape of scientific research is undergoing a profound transformation, driven by the rapidly advancing capabilities of Artificial Intelligence (AI). The concept of "AI for Science" (AI4S) has shifted from a theoretical possibility to a practical paradigm, fundamentally altering how hypotheses are generated and validated\cite{wang2023scientific, agrawal2016perspective}. Central to this revolution are Large Language Models (LLMs), which have demonstrated remarkable proficiency not only in natural language processing but also in logical reasoning and code generation\cite{chen2021evaluating, bubeck2023sparks}. While early models focused primarily on text generation, the recent integration of Chain-of-Thought (CoT) reasoning\cite{wei2022chain} and the ReAct paradigm\cite{yao2022react} has evolved LLMs from passive generators to autonomous Agents. These Agents, now equipped with modular "skills" and tool-use capabilities, can interact with external environments\cite{schick2023toolformer, xi2025rise}. This evolution promises to automate complex cognitive tasks that were previously the exclusive domain of human experts, potentially accelerating scientific discovery at an unprecedented pace\cite{boiko2023autonomous,szymanski2023autonomous,merchant2023scaling}.

Despite these technological strides, the application of AI in many specialized scientific domains faces a persistent impediment: the scarcity of high-quality data\cite{wang2022data}. While general-purpose models benefit from internet-scale datasets, domain-specific tasks, such as those in materials science, mechanics, and chemistry, rely on data that is often fragmented, unstructured, and hidden within the text and figures of academic literature\cite{fan2024openchemie}. Constructing high-fidelity datasets in these fields is notoriously labor-intensive and prohibitively expensive. Previous efforts like ChemDataExtractor\cite{swain2016chemdataextractor} and MatScholar\cite{weston2019named} made significant progress using rule-based or LSTM-based approaches, but they often struggle with complex semantic reasoning and cross-modal alignment. Manual curation remains the norm, yet it is prohibitively expensive and prone to human error\cite{wang2022dataset}. This "data bottleneck" serves as the primary obstacle preventing the widespread adoption of data-driven machine learning (ML) techniques in these critical fields\cite{zhang2018strategy}.

However, the rapid maturation of Agent frameworks and Skill-based architectures offers a novel solution to this dilemma\cite{bran2023chemcrow}. Notably, the acquisition of relevant literature in specialized fields is far from a trivial retrieval task; it entails significant complexity, requiring deep domain knowledge to discern valid experimental contexts from tangential mentions and to evaluate the methodological relevance of each source. Unlike traditional Natural Language Processing (NLP) or Optical Character Recognition (OCR) pipelines, which lack the capacity for cross-modal reasoning to resolve the semantic coupling between graphical data points and textual physical parameters (e.g., temperature $T$, stress $\sigma$)\cite{tshitoyan2019unsupervised, luo2021chartocr,masry2022chartqa, ji2023survey,lewis2020retrieval, liu2023visual}, modern Agents can emulate expert-level cognition. We posit that the current state of these frameworks has reached a level of sophistication sufficient to autonomously navigate the vast ocean of scientific literature. By defining specific skills, we empower agents to achieve end-to-end autonomy, operating entirely human-out-of-the-loop, to automatically collect relevant papers, filter for quality with professional discernment, extract distinct data points, and structure this information into queryable databases. To validate this hypothesis, we focus on the domain of material creep mechanics. We selected this field not merely as a testbed for our framework, but because it represents a quintessential class of high-dimensional, non-linear, and scarce scientific data\cite{kassner2015fundamentals}. This unique characteristic renders it an ideal touchstone for validating the physics-informed reasoning capabilities of autonomous agents, rather than merely serving as a stochastic test case. While authoritative repositories such as the NIMS Creep Data Sheets\cite{Sawada2019} provide high-fidelity benchmarks, they fundamentally rely on meticulous manual curation and standardized testing protocols. This dependence renders them inherently static and limited in scope, unable to keep pace with the vast, heterogeneous experimental results continuously emerging in academic literature. Consequently, the absence of a dynamic, large scale database containing both creep curves and their explicit constitutive expressions severely limits research progress.  On one hand, from an AI perspective, this scarcity hinders the advancement of symbolic regression tasks\cite{udrescu2020ai,karniadakis2021physics} and physics-informed neural networks (PINNs)\cite{raissi2019physics,zhang2025multitask}; without a rich repository of known functional forms and experimental data, algorithms struggle to discover or refine physical laws governing material deformation. On the other hand, for experimentalists, the absence of a centralized library makes benchmarking laborious, prohibiting the efficient comparison of new experimental results against historical data. Establishing a database in this domain would thus facilitate data-driven discovery of constitutive models and accelerate the assessment of high-temperature material performance.

Here, we demonstrate a novel workflow for the automated construction of domain-specific databases using an LLM-based Agent and Skills framework. Employing creep mechanics as a representative case study, we design a set of specialized skills that enable the agent to identify relevant literature and extract creep curves alongside their explicit mathematical expressions. Our work illustrates how the synergy between LLMs and skill-based architectures can break the data bottleneck, providing a scalable pathway for building the foundational datasets necessary for the next generation of AI4Science.

\section{Methodology}
\begin{figure*}[htbp]
    \centering
    \includegraphics[width=0.95\linewidth]{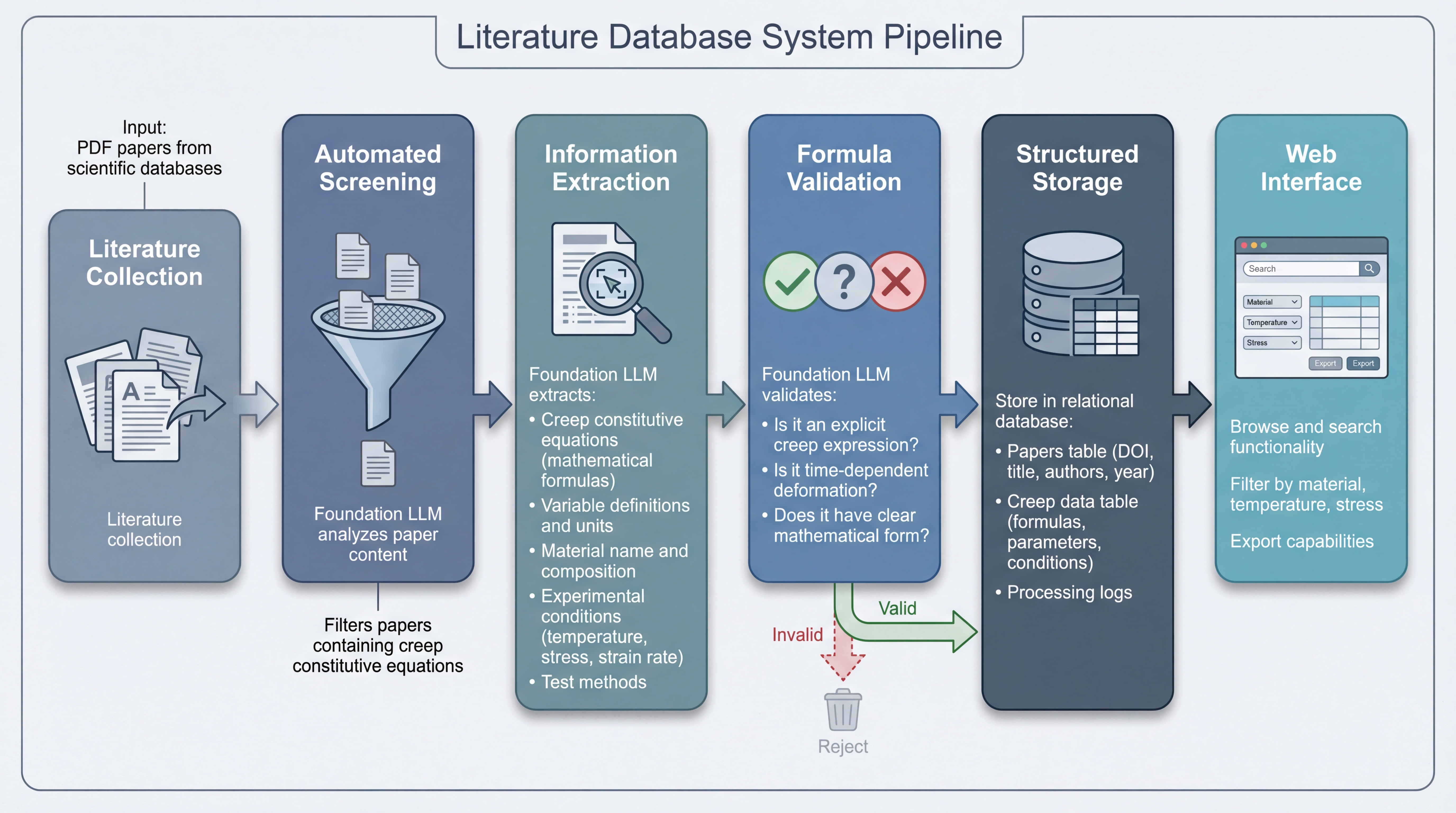}
    \caption{\textbf{Architecture of the automated literature mining pipeline.} 
    The workflow proceeds in five distinct stages: 
    (1) \textbf{Literature Collection}: Aggregation of source PDFs from scientific repositories; 
    (2) \textbf{Automated Screening}: An LLM-based filter analyzes document content to identify relevant publications containing creep constitutive models; 
    (3) \textbf{Information Extraction}: A multi-modal extraction engine parses both unstructured text (for material metadata and formulas) and graphical plots (for experimental creep trajectories); 
    (4) \textbf{Formula Validation}: A logic-based validation module checks the mathematical integrity and physical plausibility of the extracted expressions; 
    (5) \textbf{Structured Storage \& Interface}: Validated data is serialized into a relational database, accessible via a user-friendly web interface for searching, filtering, and exporting.}
    \label{fig:pipeline}
\end{figure*}

The overall architecture of the automated literature mining pipeline is illustrated in Figure \ref{fig:pipeline}. The system is designed to transform unstructured scientific PDFs into a structured, queryable database of creep behaviors and constitutive models. The workflow leverages a Foundation Large Language Model (LLM) as the central reasoning agent, orchestrating a five-stage process: (1) Literature Collection, (2) Automated Screening, (3) Multi-modal Information Extraction, (4) Formula Validation, and (5) Structured Storage. This modular design ensures that both textual metadata and graphical experimental data are processed with high fidelity and physical consistency. Specifically, the core cognitive engine of this framework is instantiated using the Qwen3-235B-A22B foundation model\cite{yang2025qwen3}, which provides the advanced logical reasoning capabilities necessary to drive these complex autonomous workflows.

\subsection{Literature Collection and Automated Screening}
The pipeline begins with the aggregation of a broad corpus of scientific literature. Potential candidate documents are sourced from major scientific repositories using a comprehensive set of keywords related to creep deformation and constitutive modeling.

Given the noise inherent in keyword-based searches, a rigorous \textbf{Automated Screening} module is employed to filter the corpus. Unlike traditional rule-based filters, this module utilizes the semantic understanding capabilities of the Foundation LLM. The agent analyzes the full text of each document to determine relevance based on specific criteria: the presence of experimental creep data and the discussion of constitutive equations. Papers that are purely theoretical (without experimental validation) or unrelated to time-dependent deformation are automatically rejected, ensuring that downstream processing is focused only on high-value content.

\subsection{Multi-modal Information Extraction}
The core of the system is the extraction engine, which handles the heterogeneous nature of scientific documents by processing two distinct modalities:
\begin{itemize}
    \item \textbf{Semantic extraction (Text \& Formulas):}The agent scans the document to identify and extract critical metadata, including material designations (e.g., chemical composition, standard names), experimental conditions (temperature $T$, stress $\sigma$), and explicit mathematical expressions of creep constitutive laws. Special attention is paid to variable definitions and unit standardization to ensure consistency across the database.
    \item \textbf{Visual extraction (Graphical Data):} Unlike indiscriminate digitization, the agent is designed to prioritize representative experimental trajectories that can be strictly mapped to the constitutive models explicitly discussed in the text. Complementing the textual analysis, the agent identifies and processes experimental plots (e.g., Strain vs. Time curves) while filtering out supplementary or ambiguous data series. It detects coordinate systems, identifies axis scales, and digitizes the data points from relevant trajectories. This capability allows the system to capture raw experimental data that is often absent from the text but crucial for model validation.
\end{itemize}

\subsection{Physics-Informed Validation Protocol}

To mitigate the risk of "hallucinations" common in generative AI models, a Formula Validation module is integrated into the workflow as a physics-informed logic gate. This module rigorously evaluates each extracted entry against a triad of criteria to ensure scientific rigor. Specifically, the agent verifies whether the extracted content constitutes a complete, explicit mathematical equation rather than a mere descriptive fragment; checks for physical relevance to ensure the formula describes time-dependent deformation consistent with established creep mechanics; and confirms mathematical integrity by ensuring all variables are clearly defined and unit-consistent. Only entries that successfully pass this comprehensive validation check are marked as valid and proceed to the database, while non-compliant entries are automatically flagged for rejection or manual review.

\subsection{Database Construction and Web Interface}

\begin{figure*}[htbp]
    \centering
    \includegraphics[width=0.95\linewidth]{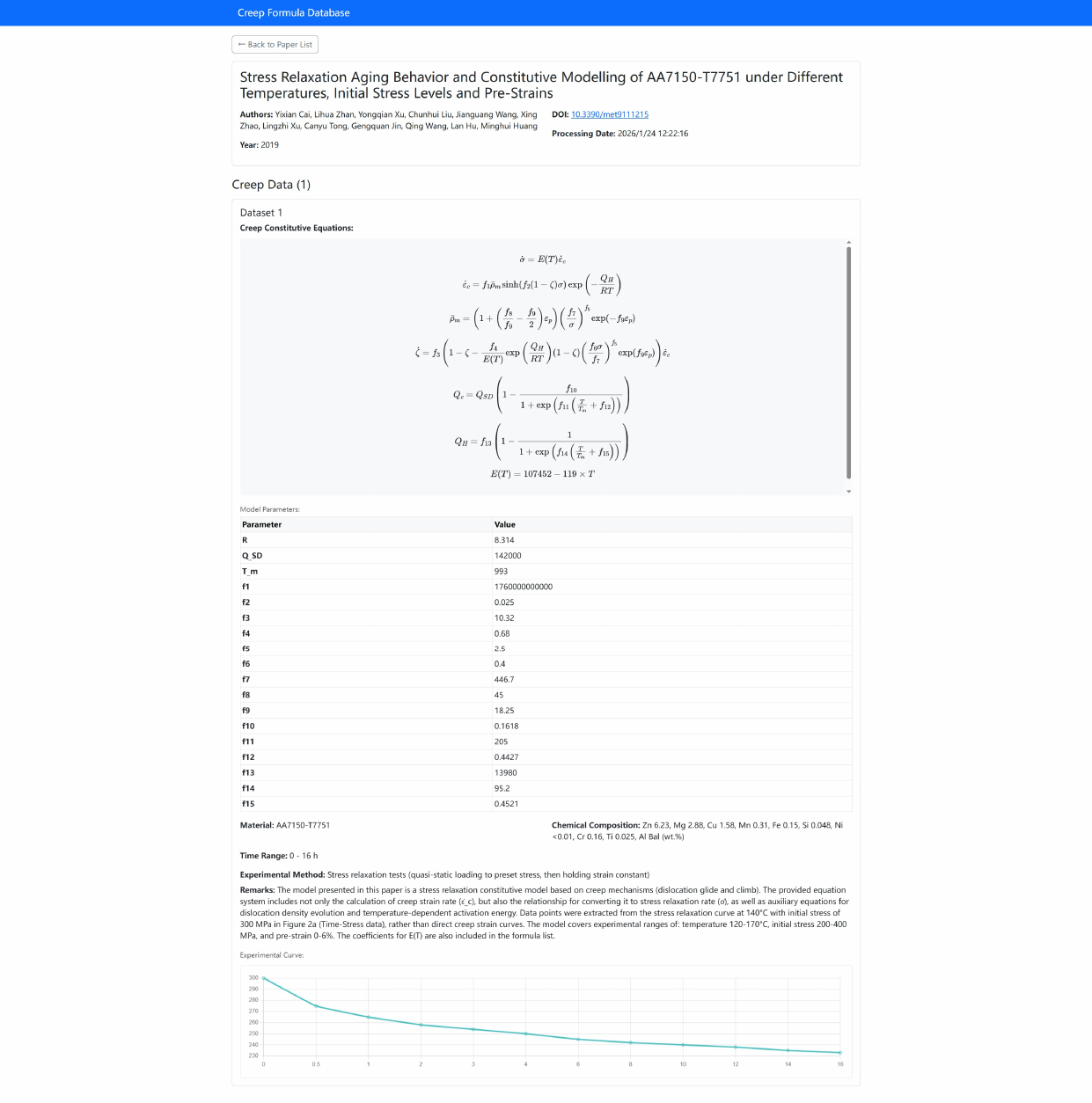}
    \caption{\textbf{Interactive front-end interface of the AI constructed Creep Database.} 
    The web-based platform acts as a bridge between the structured SQL backend and end users, facilitating efficient data retrieval. Key features include: 
    (1) \textbf{Granular Filtering} (Left Panel): Users can query the database using multidimensional constraints, such as material composition, temperature ranges, and stress levels; 
    (2) \textbf{Dynamic Visualization} (Right Panel): Selected datasets are instantly rendered into strain time plots, allowing for rapid visual comparison of creep behaviors across different experimental conditions; 
    (3) \textbf{Data Export}: Validated data points and constitutive parameters can be exported in standardized formats for downstream machine learning applications.}
    \label{fig:web_interface}
\end{figure*}

Verified data is serialized into a relational database structure comprising two primary entities: a Papers Table for bibliographic metadata (DOI, authors, year) and a Creep Data Table for the extracted physics parameters and experimental conditions. Crucially, the preservation of the Digital Object Identifier (DOI) is implemented to guarantee data provenance, ensuring that every extracted data point is strictly traceable to its original source publication. To facilitate accessibility, a user-friendly Web Interface was developed on top of the database. As depicted in Figure \ref{fig:web_interface}, the interface provides functionality for browsing and searching. Users can filter the database by material type, temperature range, or stress levels, and export the structured datasets for further analysis. This end-to-end implementation ensures that the extracted knowledge is not only stored but readily available for data-driven AI4S discovery.

\subsection{Skill-Based Cognitive Architecture and Constraint Mechanisms}
To execute the aforementioned workflow with high autonomy and reliability, we engineered a modular ``Skill-Based'' cognitive architecture. This framework departs from monolithic agent designs by decomposing complex reasoning tasks into specialized, executable units defined as Skills. Formally, we model an Agent Skill ($S$) as a cognitive tuple $S = \{I, T, C\}$, where:
\begin{itemize}
    \item $I$ (Instruction Injection): Represents domain-specific system prompts that ``prime'' the LLM with expert knowledge  while explicitly masking irrelevant general knowledge to reduce noise.
    \item $T$ (Scoped Tools): A restricted set of executable functions available exclusively to that specific skill. This action space pruning minimizes the probability of hallucinated or erroneous tool calls.
    \item $C$ (Hard Constraints): A mandatory output schema that enforces constrained generation, ensuring machine-actionability of the outputs.
\end{itemize}
As illustrated in Table~\ref{tab:skill_map}, this architecture allows the agent to dynamically instantiate specialized personas---ranging from a ``Bibliographer'' to a ``Physicist''---at different stages of the pipeline, enforcing strict logic gates to improve data fidelity.

\begin{table}[htbp]
\centering
\caption{Mapping of pipeline stages to agent skills and constraint logic.}
\label{tab:skill_map}
\renewcommand{\arraystretch}{1.8}
\small
\begin{tabular}{>{\centering\arraybackslash}m{1.8cm} >{\centering\arraybackslash}m{2.5cm} >{\centering\arraybackslash}m{2.5cm} m{6.5cm}}
\hline
\textbf{Stage} & \textbf{Agent Skill} & \textbf{Scoped Tools($T$)} & \textbf{Constraint Mechanism ($C$)} \\
\hline
Collection & Bibliographic Navigator & RAG-based Retrieval & \textbf{Semantic Disambiguation}: Transforms ambiguous NL descriptors into precise, syntax-compliant Boolean logic; expands implied keywords (e.g., "superalloy" $\rightarrow$ "Ni-based OR Co-based") to maximize recall. \\
Screening & Domain Filter & Zero-shot Classification & \textbf{Logical Gating}: Implements a strict conjunctive filter ($Pass \iff Has\_Data \land Has\_Equation$); automatically rejects partial matches (e.g., pure simulation studies) regardless of keyword relevance. \\
Extraction & MultiModal Parser & Vision-Language Alignment & \textbf{Ontological \& Geometric Alignment}: Enforces a unified variable schema to resolve nomenclatural heterogeneity, while imposing strict topological constraints (e.g., monotonicity) to guarantee the structural integrity of digitized trajectories. \\
Validation & Physics Guardrail & Symbolic Reasoning & \textbf{Physics-Informed Integrity}: Mandates adherence to dimensional homogeneity laws and enforces a rigorous statistical threshold ($R^2 > 0.9$) for cross-modal alignment between textual parameters and visual data. \\
Storage & Data Serializer & Structured Serialization & \textbf{Schema Compliance \& Provenance}: Enforces strict data typing to ensure machine-readability, while embedding immutable metadata (e.g., source DOI, visual evidence links) to establish a verifiable audit trail for every entry. \\
\hline
\end{tabular}
\end{table}

\subsection{Quality Assessment of the Automated Workflow}
To evaluate the efficacy of the automated screening module, a quantitative assessment was performed against a human-verified ground truth. A validation subset consisting of 20 randomly selected documents was established from the initial corpus. We manually reviewed these documents to categorize them as either "Relevant" (containing valid creep experimental data) or "Irrelevant".
Given that this process functions as an information retrieval task, we adopted standard evaluation metrics used in Retrieval-Augmented Generation (RAG) systems to quantify performance: Precision, Recall, F1-Score, and Accuracy. To ensure clarity for readers across disciplines, we define these metrics based on the following classification outcomes:
\begin{itemize}
\item \textbf{True Positive ($TP$)}: The Agent correctly identified a relevant paper containing creep data.
\item \textbf{False Positive ($FP$)}: The Agent incorrectly classified an irrelevant paper as relevant (noise).
\item \textbf{True Negative ($TN$)}: The Agent correctly rejected an irrelevant paper.
\item \textbf{False Negative ($FN$)}: The Agent failed to identify a relevant paper (missed data).
\end{itemize}
Based on these counts, the metrics are calculated as follows:
\textbf{Precision} measures the reliability of the agent's selection, indicating the percentage of retrieved documents that are actually relevant:
\begin{equation}
\text{Precision} = \frac{TP}{TP + FP}
\end{equation}
\textbf{Recall} measures the completeness of the search, indicating the percentage of all existing relevant documents that the agent successfully found:
\begin{equation}
\text{Recall} = \frac{TP}{TP + FN}
\end{equation}
The \textbf{F1-Score} provides a harmonic mean of Precision and Recall, offering a balanced view of the agent's robustness:
\begin{equation}
\text{F1-Score} = 2 \times \frac{\text{Precision} \times \text{Recall}}{\text{Precision} + \text{Recall}}
\end{equation}
Finally, \textbf{Accuracy} represents the overall correctness of the classifier across the entire validation set:
\begin{equation}
\text{Accuracy} = \frac{TP + TN}{TP + TN + FP + FN}
\end{equation}
These metrics provide a comprehensive framework to benchmark the agent's ability to maximize data acquisition while minimizing the inclusion of irrelevant literature.

\section{Results}\label{sec2}
\begin{figure}[htbp]
    \centering
    \includegraphics[width=0.95\linewidth]{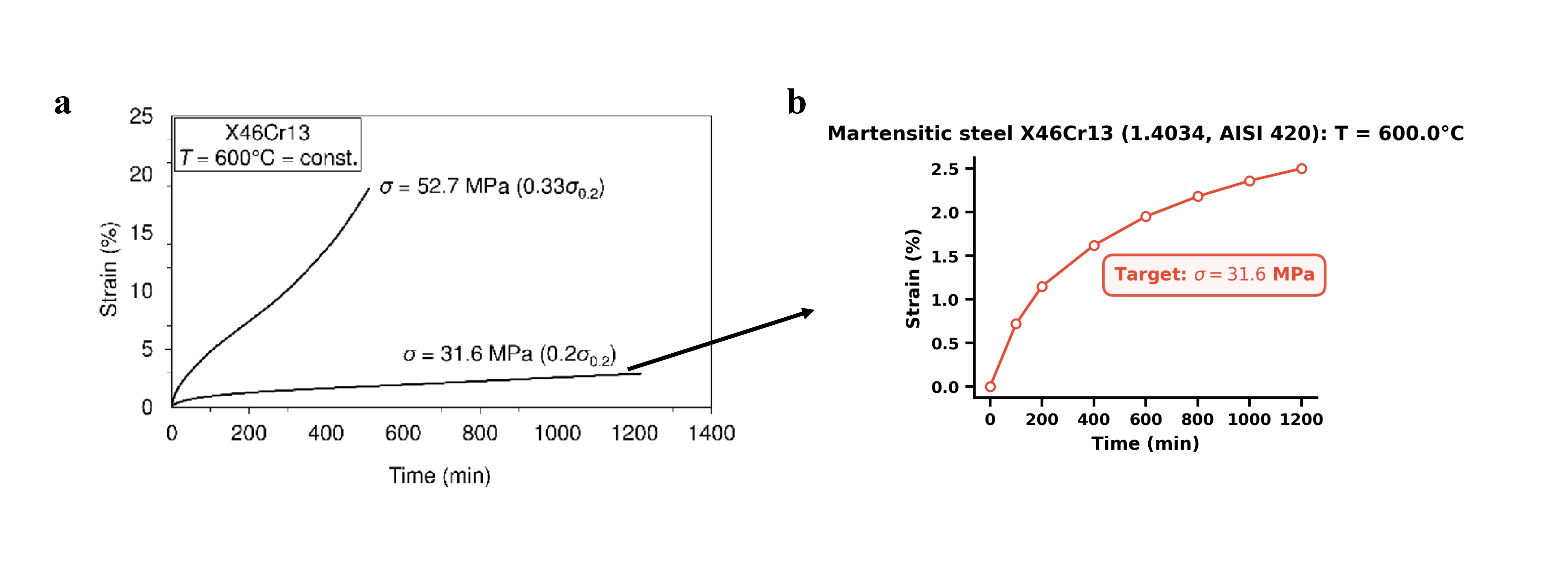}
    \caption{\textbf{Demonstration of selective curve extraction from complex multi-variable plots.} 
    (a) Original source image (Material: X46Cr13, T=600$^\circ$C) containing multiple creep trajectories under different stress levels\cite{ma10040388}. 
    (b) The digitized result reconstructed from the agent's output. The agent successfully disentangled the visual information, accurately isolating the specific curve corresponding to the target condition ($\sigma = 31.6$ MPa) while ignoring the extraneous data series (52.7 MPa). Note that the high fidelity of the extracted data points preserves the underlying creep trend.}
    \label{fig2}
\end{figure}

To ensure the reliability of the constructed database, we implemented a rigorous "human-in-the-loop" verification protocol. Following the automated extraction phase, a random subset of the processed literature was subjected to manual inspection. Out of the 243 processed documents, the agent achieved a verified success rate exceeding 90\% in correctly identifying and digitizing target creep curves. This high metric is particularly significant given the heterogeneous nature of the source documents, which range from high-resolution modern digital PDFs to scanned legacy publications spanning several decades, often characterized by varying clarity and rasterization artifacts.
Figure \ref{fig2} presents a qualitative comparison that highlights the agent's robustness in handling complex extraction scenarios. A major challenge in automated literature mining is the high density of visual information; a single publication typically contains various experimental visualizations—ranging from tensile stress-strain curves to microstructural micrographs—that must be distinguished from the target creep data. Furthermore, even within a identified creep plot, multiple trajectories often coexist, representing different stress or temperature conditions (as shown in Figure \ref{fig2}a, where curves for 52.7 MPa and 31.6 MPa overlap). As illustrated in the reconstructed result (Figure \ref{fig2}b), the agent successfully performed semantic disentanglement: it not only filtered out graphical artifacts like grid lines and background noise but also accurately isolated the specific trajectory corresponding to the target experimental condition ($\sigma = 31.6$ MPa), effectively ignoring the extraneous data series. This demonstrates a context-aware extraction capability that transcends simple visual segmentation.
Beyond the graphical data, the semantic accuracy of the extracted metadata—specifically the experimental conditions (temperature, stress) and material identifiers—was also validated. The consistency of these textual extractions is evidenced by the logical clustering observed in the previously discussed parameter space, where no distinct physical outliers (e.g., ice appearing in high-temperature clusters) were detected. This dual validation of both visual (curve) and semantic (text) information confirms that the automated workflow is sufficiently precise to serve as a foundation for data-driven scientific inquiry.

\subsection{Overview and Composition of the Constructed Database}\label{subsec1}

To evaluate the efficacy of the proposed Agent-Skills framework, we deployed the system to perform an automated extraction task on a collection of unstructured academic literature. The agent processed 243 publications and successfully constructed 353 high-fidelity 'Equation-Data' pairs. While a single publication may contain multiple graphical datasets, our framework focuses on isolating the principal trajectories used by the original authors to derive or validate their constitutive models. This rigorous filtering resulted in a focused dataset where each curve is accompanied by its explicit mathematical parameters, prioritizing physical interpretability over raw data volume. A subsequent manual verification of the extracted data revealed a high fidelity in the automated workflow, with a validated success rate exceeding 90\%. This high precision underscores the robustness of the defined skills in handling the heterogeneous formatting and nomenclature found across different journals and decades of research.

\begin{figure}[htbp]
\centering
\includegraphics[width=0.9\textwidth]{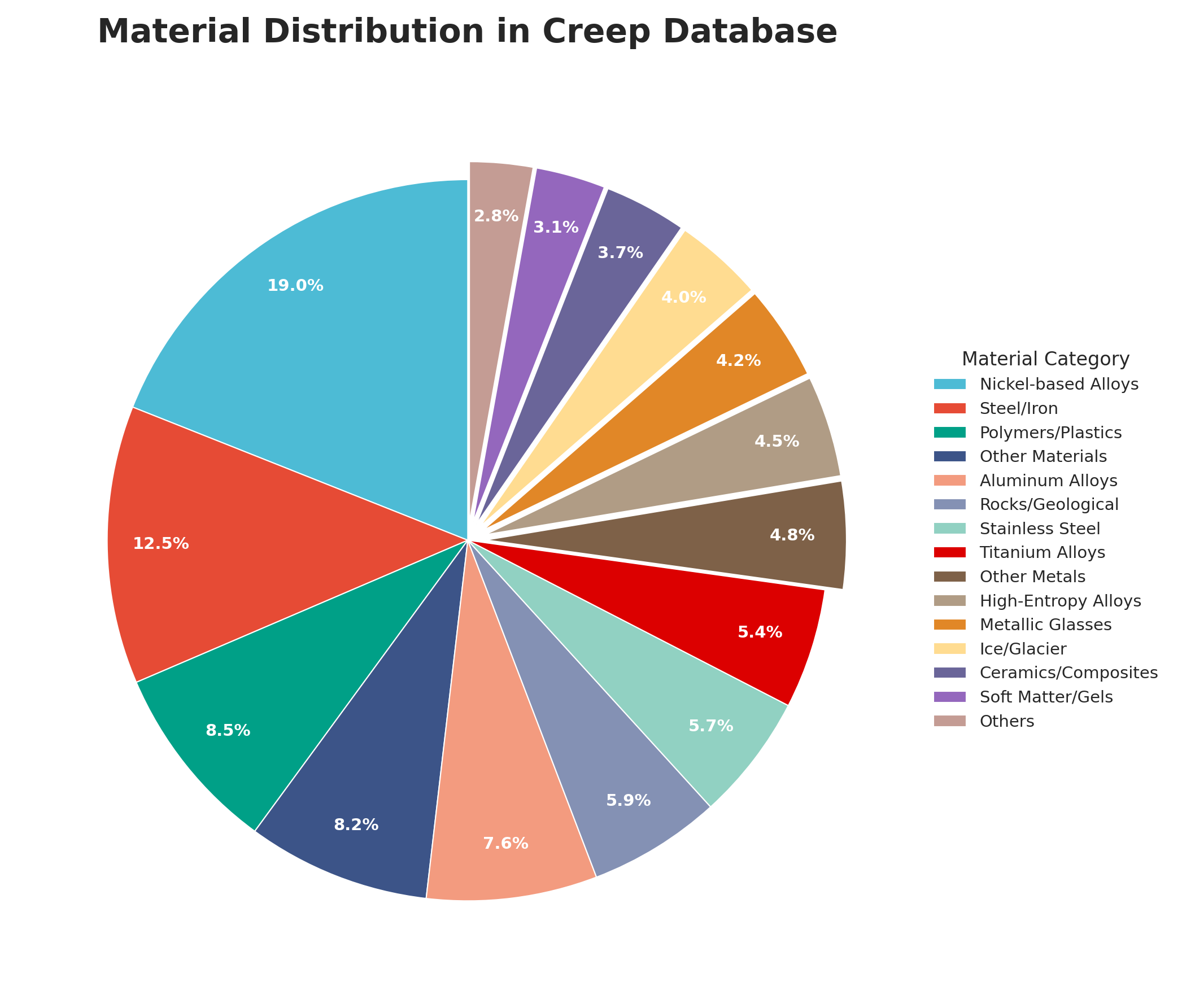}
\caption{\textbf{Diversity of material composition within the constructed creep database.} The sector chart illustrates the distribution of material categories across the 353 extracted creep curves. The dataset demonstrates broad domain coverage, encompassing critical industrial materials (e.g., Nickel-based alloys, 19.0\%; Steel/Iron, 12.5\%) as well as non-metallic systems (e.g., Polymers, 8.5\%; Ice/Glaciers, 4.0\%). This diversity confirms the agent's semantic ability to identify relevant ``creep'' behaviors across heterogeneous material science sub-domains.}\label{fig3}
\end{figure}

The resulting database exhibits a remarkable diversity in material composition, demonstrating the agent's ability to generalize across various scientific sub-domains. As illustrated in Figure \ref{fig3} (Material Distribution), the dataset is predominantly composed of high-performance engineering materials, with Nickel-based alloys (19.0\%) and Steel/Iron (12.5\%) constituting the largest shares. These are critical materials for high-temperature applications, reflecting the literature's focus on industrial relevance. Notably, the agent also successfully retrieved data for less conventional categories, including Polymers/Plastics (8.5\%), Rocks/Geological materials (5.9\%), and even Ice/Glaciers (4.0\%). This inclusion of diverse material classes—ranging from soft matter to geological composites—validates that the agent is not over-fitted to specific metallic keywords but possesses a semantic understanding of the "creep" phenomenon broadly defined.

\begin{figure}[htbp]
\centering
\includegraphics[width=0.9\textwidth]{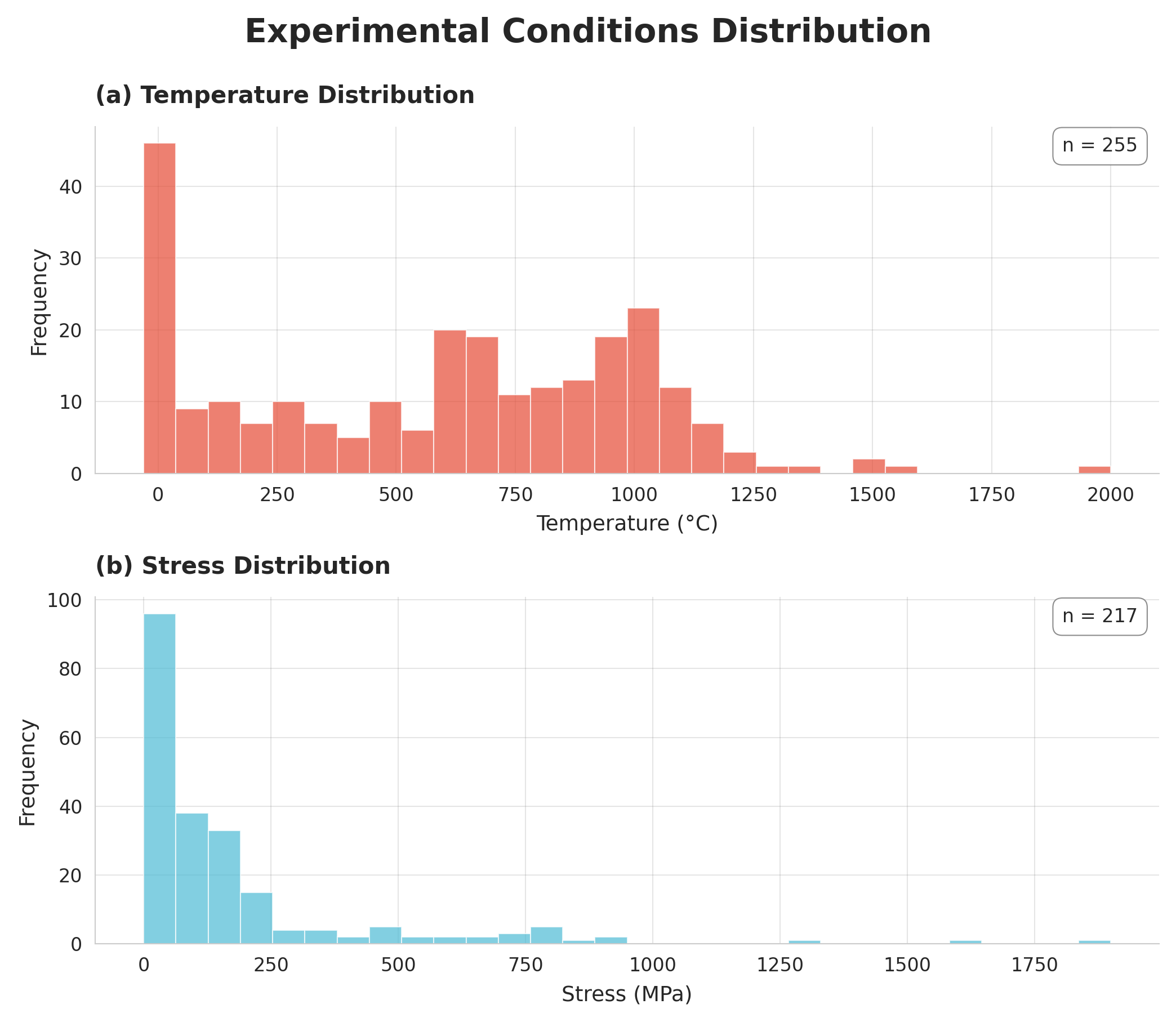}
\caption{\textbf{Statistical distribution of experimental conditions extracted from the literature.} (a) Frequency distribution of experimental temperatures ($^\circ$C), spanning from cryogenic environments to ultra-high temperature regimes ($>1000^\circ$C). (b) Frequency distribution of applied stress levels (MPa). The broad range of operating conditions indicates that the database captures a comprehensive snapshot of material performance boundaries documented in historical literature.}\label{fig4}
\end{figure}

\begin{figure}[htbp]
\centering
\includegraphics[width=0.9\textwidth]{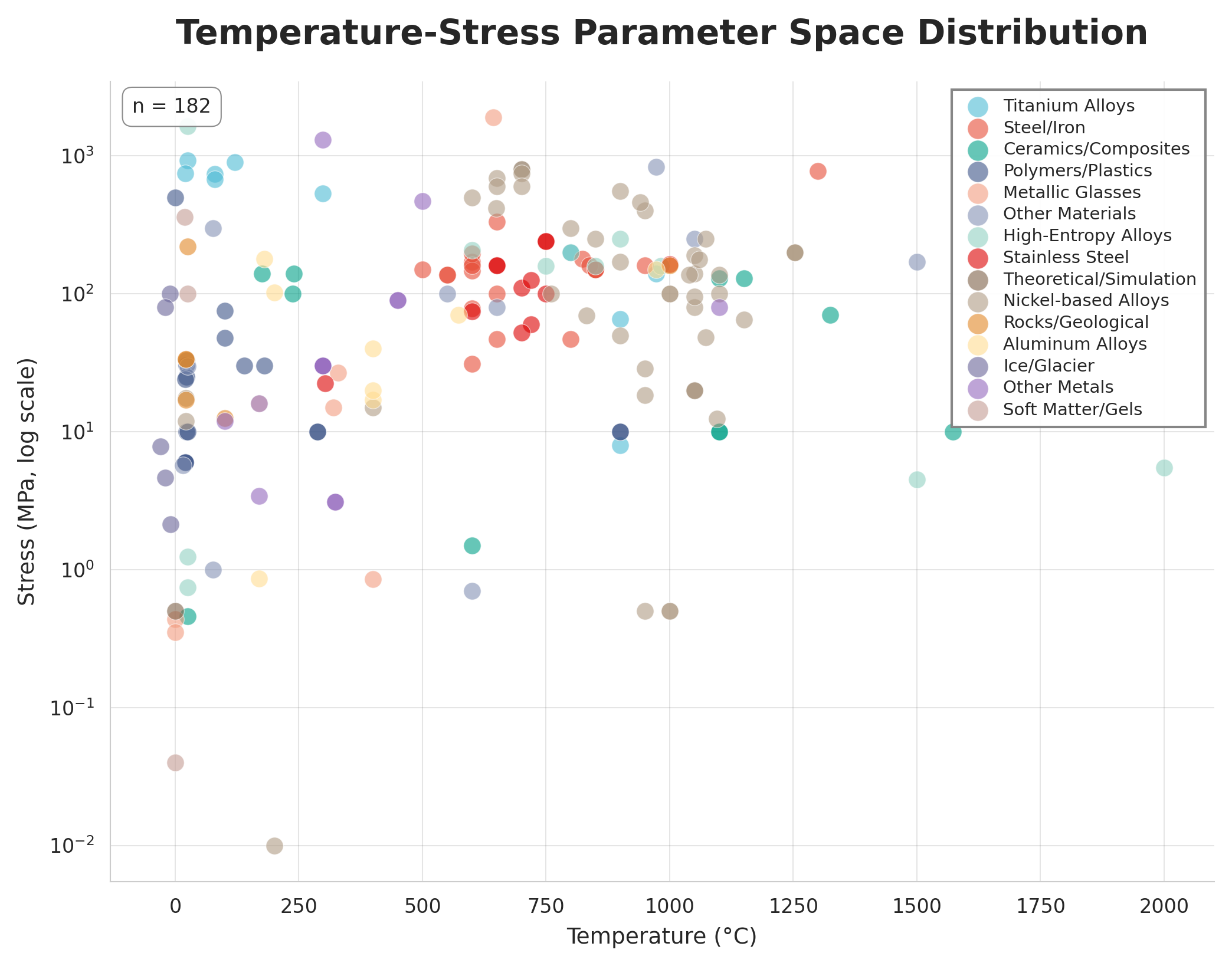}
\caption{\textbf{Visualization of the Temperature-Stress parameter space.} A scatter plot mapping the experimental conditions for all extracted curves ($n=353$), color-coded by material category. The visualization reveals distinct physical clusters corresponding to material capabilities—such as the high-temperature/high-stress dominance of superalloys versus the low-energy regimes of soft matter—demonstrating the physical consistency and logical distribution of the automatedly constructed dataset.}\label{fig5}
\end{figure}

Furthermore, the database covers a comprehensive range of experimental conditions, providing a rich parameter space for future modeling. Figure \ref{fig4} visualizes the distribution of key physical parameters: temperature and stress. The temperature distribution (Figure \ref{fig4}a) spans from cryogenic and room-temperature environments (typical for polymers and ice) to extreme conditions exceeding 1000°C (characteristic of superalloys and ceramics). Similarly, the stress levels (Figure \ref{fig4}b) range from low-magnitude geological pressures to high-stress engineering loads (>1000 MPa). The interaction between these variables is mapped in the Temperature-Stress parameter space shown in Figure \ref{fig5}. This scatter plot reveals distinct clusters corresponding to different material capabilities—for instance, the high-temperature/high-stress region occupied by superalloys versus the low-temperature/low-stress region of soft matter. The successful population of this broad physical landscape confirms that the constructed database is not only sizable but also physically representative, effectively capturing the constitutive behaviors of materials across multiple scales.

\subsection{Verification of Physical Consistency via Cross-Modal Alignment}
\begin{figure}[htbp]
    \centering
    \includegraphics[width=0.95\linewidth]{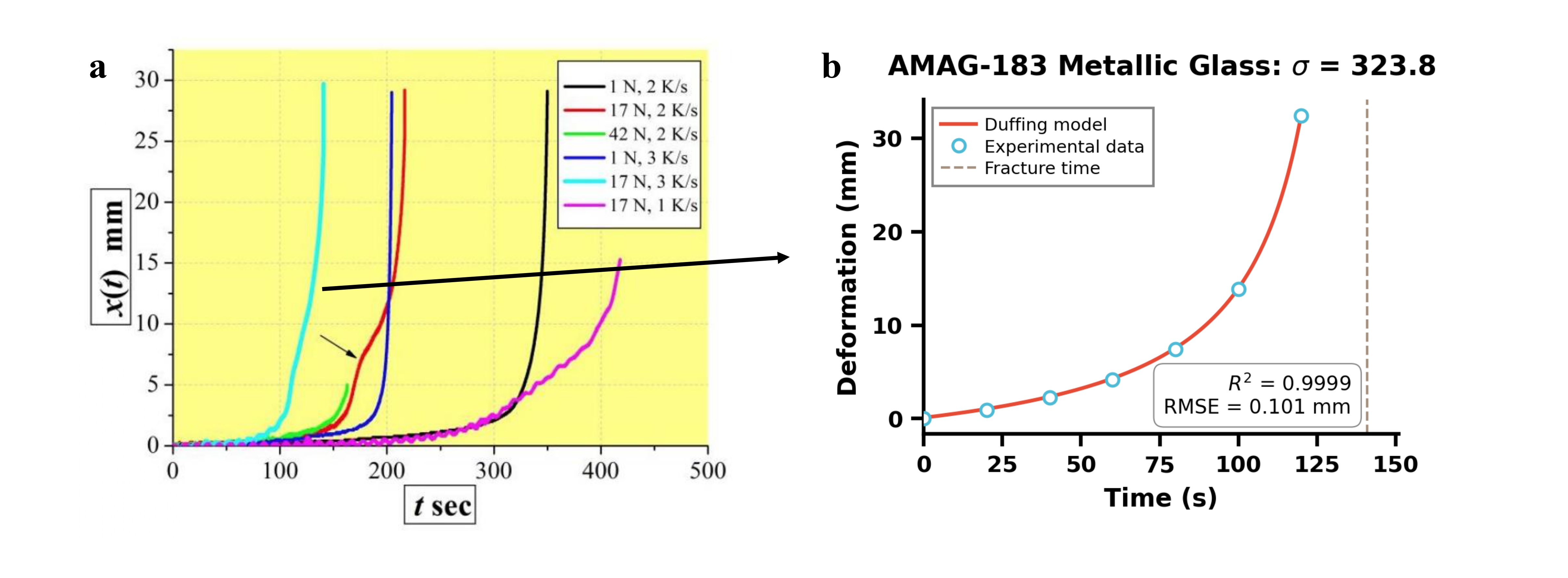}
    \caption{\textbf{Cross-modal verification of physical consistency for complex non-linear behavior.} 
    (a) Original experimental plot for AMAG-183 metallic glass, characterized by a non-standard background and multiple overlapping series\cite{berezner2025novel}. 
    (b) Validation of the extracted data against the theoretical physics model. The symbols (blue circles) represent the data digitized from the image, while the solid line (red) represents the trajectory calculated using the \textit{Duffing model} parameters extracted from the text. The exceptional correlation ($R^2 = 0.9999$) demonstrates the agent's ability to correctly map textual mathematical descriptions to their corresponding visual representations, validating the physical integrity of the constructed database.}
    \label{fig:consistency}
\end{figure}

Beyond graphical accuracy, a critical requirement for a scientific database is the preservation of physical consistency between the extracted raw data and the theoretical models described in the source text. Scientific publications often exhibit information redundancy, presenting experimental results in graphical forms while simultaneously reporting the fitted constitutive parameters in the accompanying text. This offers a unique opportunity for cross-modal validation.
To demonstrate this, we analyzed a representative case study on AMAG-183 metallic glass (Figure \ref{fig:consistency}). The agent independently executed two tasks: (1) digitizing the raw deformation-time trajectory from a complex plot characterized by colored backgrounds and multiple data series (Figure \ref{fig:consistency}a); and (2) extracting the mathematical modeling parameters from the document's discussion section. In this specific case, the source literature described the deformation behavior using a non-linear Duffing oscillator model to capture the chaotic dynamics of shear band evolution.
Figure \ref{fig:consistency}b visualizes the alignment between these two modalities. The open circles represent the visually digitized experimental data, while the solid red line represents the theoretical trajectory reconstructed by substituting the textually extracted parameters into the governing differential equation. As quantified in the inset, the theoretically reconstructed model shows exceptional agreement with the digitized data points, achieving a coefficient of determination ($R^2$) of 0.9999. This near-perfect synchronization confirms that the agent effectively bridged the semantic gap between unstructured textual parameters and rasterized visual data, ensuring the database is physically self-consistent even for complex non-linear material behaviors. This rigorous cross-modal verification protocol implicitly acts as a quality filter; while it may limit the total number of extracted curves per publication compared to indiscriminate extraction, it guarantees that every entry in the database possesses a verifiable high-fidelity link between experimental data and physical theory.

\section{Conclusion}
In this study, we presented a novel, autonomous framework capable of transforming the fragmented landscape of scientific literature into structured, machine-actionable knowledge. By synergizing the reasoning capabilities of Foundation LLMs with precise visual extraction skills, we successfully breached the "data bottleneck" that has long impeded data-driven research in mechanics and materials science.
The construction of the creep mechanics database served as a rigorous testbed for this approach. Our results demonstrate that the agent transcends traditional OCR-based limitations, exhibiting the semantic intelligence necessary to disentangle complex experimental plots and validate data against physical laws. The achievement of a near-perfect correlation ($R^2 = 0.9999$) between digitized curves and theoretical models in our case study highlights the system's ability to perform cross-modal reasoning—effectively bridging the gap between visual observations and mathematical descriptions.
While this work focused on creep deformation, the underlying architecture is domain-agnostic. The modular nature of the "Agent-Skill" framework allows for rapid adaptation to other specialized fields, from battery cycling performance to catalytic reaction kinetics. As LLMs continue to evolve, we envision such autonomous data agents becoming a standard component of the scientific infrastructure, liberating researchers from tedious data curation and accelerating the loop of hypothesis generation and discovery in the era of AI-driven science.

\bibliographystyle{unsrt}  


\bibliography{references}

\end{document}